\documentclass[apj]{emulateapj}
\usepackage{rotating,psfig,natbib}

\def\simgr{\,\hbox{\hbox{$ > $}\kern -0.8em \lower 1.0ex\hbox{$\sim$}}\,}
\def\simle{\,\hbox{\hbox{$ < $}\kern -0.8em \lower 1.0ex\hbox{$\sim$}}\,}

\begin{document}
\bibliographystyle{apj}

\slugcomment{Received 2007 October 26; accepted 2008 February 5}

\title{On the Presence of Water and Global Circulation in the Transiting Planet HD 189733b}

\author{Travis S. Barman}
\affil{Lowell Observatory, 1400 W. Mars Hill Rd., Flagstaff, AZ 86001\\ 
        Email: {\tt barman@lowell.edu}}

\begin{abstract}
Detailed models are compared to recent infrared observations of the nearby
extrasolar planet, HD 189733b. It is demonstrated that atmospheric water is
present and that the planet's day side has a non-isothermal structure down to
gas pressures of $\sim$ 0.1 bars.  Furthermore, model spectra with different
amounts of CO are compared to the observations and an atmosphere absent of CO
is excluded at roughly 2-sigma.  Constraining the CO concentration beyond that
is unfortunately not possible with the current $Spitzer$ photometry.  However,
radically enhanced (or depleted) metal abundances are unlikely and the basic
composition of this planet is probably similar to that of its host star.  When
combined with $Spitzer$ observations, a recent ground-based upper limit for the
$K$-band day side flux allows one to estimate the day-to-night energy
redistribution efficiency to be $\sim$ 43\%.
\end{abstract}


\section{Introduction}
Hundreds of extrasolar planets have now been discovered along with over a dozen
transiting systems.  Despite the rapid pace of discovery, many important
questions remain unanswered, especially concerning the transiting giant planets
(hot-Jupiters). How efficiently is incident stellar energy absorbed by the day
side redistributed to the night side?  Do hot-Jupiters have atmospheres with
enhanced metal abundances relative their host star?  And under what conditions
(and depths) are the day side atmospheric temperature structures isothermal?
Fortunately, for the handful of nearby transiting planets, direct flux
measurements have been obtained using the {\em Spitzer Space Telescope}.
Fluxes provide a direct look into a planet's depth-dependent atmospheric
conditions and perhaps hold the answers to some of these questions.

HD 189733b, is the closest known jovian-mass transiting planet
\cite[]{Bouchy05}, and flux measurements of its day side are available across a
wide range of infrared wavelengths along with a recent measurement of the
planet's night side at 8 $\micron$ \cite[]{Deming06, Grillmair07, Charb08,
Knutson07b}.  HD 189733b is expected to have photospheric temperatures in the
1000 to 1500K range on its day side maintained by the large amount of incident
stellar flux.  At just 20 pc away from Earth, this planet is a valuable test
case for atmospheric dynamics and chemical enrichment by accretion of
extrasolar comets and asteroids.

In this Letter atmospheric models are compared to recent IR observations of HD
189733b to infer some of its basic photospheric properties.  The concentrations
of important molecules (e.g., water and CO) are discussed along with the
day-to-night energy redistribution by depth-dependent horizontal circulations.

\section{Eclipse Depths and Spectra}

As a transiting planet passes behind its host star a small but measurable drop
in infrared flux occurs as the planet's day side contribution disappears.  A
few hours later, the planet reappears from behind the host star and the total
system flux returns to its previous value.  The change in flux during this
secondary eclipse provides a precise measurement of the day side planet-star
flux density ratio ($\epsilon_{\lambda} = \left[\frac{R_{p}^2 F_{p}}{R_{\star}^2
F_{\star}}\right]_\lambda$, where $R_{p,\star}$ are the planet and star radii)
at a given wavelength.  Since even heavily irradiated planets are much cooler
than their host star, their peak flux emerges at a significantly longer
wavelength ($\lambda_{p}^{max}$) than it does for the host star causing
$\epsilon_{\lambda}$ to rise toward longer wavelengths and making secondary
eclipse observations more practical at $\lambda \simgr \lambda_{p}^{max}$.
Regardless of the detailed absorption features in the two spectra,
$\epsilon_{\lambda}$ falls rapidly from the infrared to the optical as a result
of the steep rise of the stellar spectrum and, consequently, the deepest
eclipse measurements made with $Spitzer$ are generally where the planet flux
density is small.

\begin{figure}
\hspace{-1cm}\includegraphics[width=10cm]{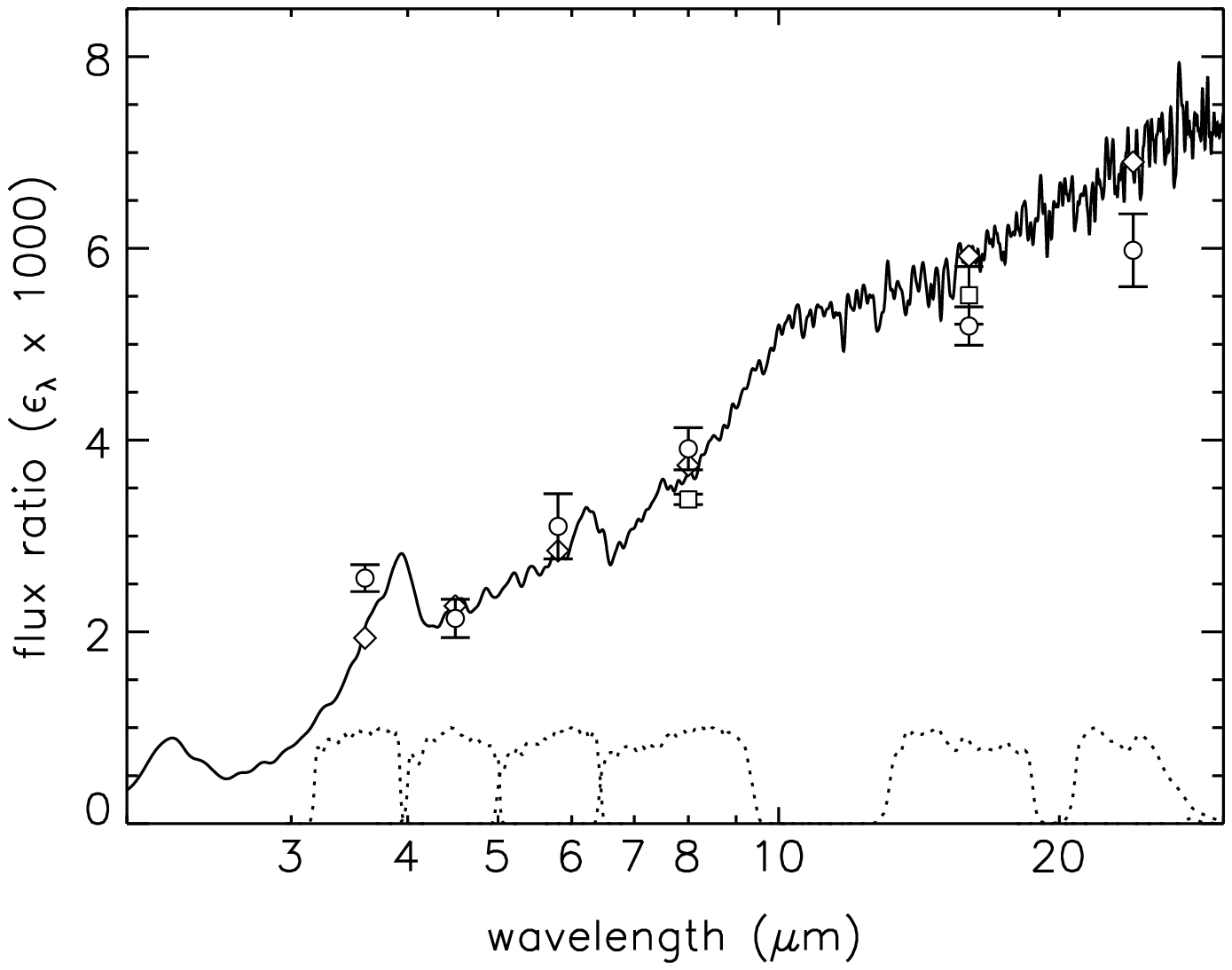}
\caption{Model planet-star flux density ratios ($\epsilon_{\lambda}$) assuming
no energy is redistributed to the night side. White circles are the
\cite{Charb08} $Spitzer$ secondary eclipse measurements with 1-$\sigma$
error-bars. White boxes are additional 8.0 and 16 $\micron$ eclipse values
\cite[]{Deming06,Knutson07b}.  White diamonds are the model values integrated
across the $Spitzer$ instrumental response curves, indicated by dotted lines.
\label{epsfig}}
\end{figure}

Secondary eclipses of HD 189733 have been observed photometrically with
$Spitzer$ at 3.6, 4.5, 5.8, 8.0, 16 and 24 $\micron$ \cite[]{Deming06, Charb08,
Knutson07b}.  A secondary eclipse has also been observed spectroscopically from
7.5 to 15 $\micron$ \cite[]{Grillmair07}.  Figure \ref{epsfig} shows all of the
day side $Spitzer$ secondary eclipse photometry for HD 189733 compared to a
synthetic $\epsilon_{\lambda}$ spectrum produced by dividing a synthetic planet
spectrum by a synthetic stellar spectrum (optimized for the K-type stellar
host).  The $\epsilon_{\lambda}$ spectrum has also been scaled by the
planet-star surface areas determined from transit observations
\cite[]{Bouchy05}.  

Given that $\epsilon_{\lambda}$ emphasizes the faintest portions of the
planet's spectrum, comparing actual planet fluxes to model spectra is more
revealing.  The observed fluxes shown in Fig. \ref{ffig} were calculated using
the out-of-eclipse stellar fluxes and the definition of $\epsilon_{\lambda}$
above.  As can be seen, the planet's spectrum peaks at $\sim 3.6 \micron$ --
far from the deepest $Spitzer$ eclipse depth.  Also, while $\epsilon_{\lambda}$
is nearly flat from 3.6 to 4.5 $\micron$, the individual planet spectrum is
quite steep across this wavelength range.

Overall, the observations agree fairly well with the model shown in Fig.
\ref{ffig} (same model in Fig. \ref{epsfig}).  The standard model includes a
solar abundance composition in chemical equilibrium and irradiation following
the methods described in \cite{Barman01,Barman05}.  Furthermore, this model
assumes no redistribution of absorbed stellar flux from the day side to the
night side and, hence, implies a very cold night side.  Interestingly, zero
energy redistribution to the night side is at odds with recent 8 $\mu$m
observations indicating a warm night side photosphere \cite[]{Knutson07b}.
Also, \cite{Burrows06} found that their full-redistribution model (50\% of the
incident energy carried to the night side) is in good agreement with the
\cite{Deming06} 16 $\mu$m $Spitzer$ observation.  However, the Burrows et al.
model predicts stronger absorption across the central two IRAC bands than
is observed.  

\section{Evidence for Water and CO}

For the temperatures and pressures expected in the atmospheres of irradiated
giant planets, models predict that H$_2$O should be the third or fourth most
abundant gas-phase molecule.  With its broad absorption features across the
IR, H$_2$O plays an important role in sculpting a giant planet's emergent
spectrum and temperature structure.  Therefore, measuring the water abundance
is an important step toward understanding the planet's overall chemistry and
structure.  

Given that basic predictions of hot-Jupiters include a copious water supply,
the first spectra of hot-Jupiters were somewhat surprising in that no water
absorption was detected.  These $\epsilon_{\lambda}$ spectra turned out to be
fairly flat from 7.5 to 11 $\micron$ which was interpreted as an absence of
water absorption in the planet's spectrum \cite[]{Grillmair07,Richardson07}.  A
few months after the IRS spectra were published, evidence for water absorption
in the atmosphere of HD 209458b was identified \cite[]{Barman07} and a similar
claim followed for HD 189733b \cite[]{Tinetti07b}.  These water detections
relied on observations during primary transit, and thus relate to different
regions of the planets than the secondary eclipse spectra (and photometry).

\begin{figure}
\hspace{-0.7cm}\includegraphics[width=9.5cm]{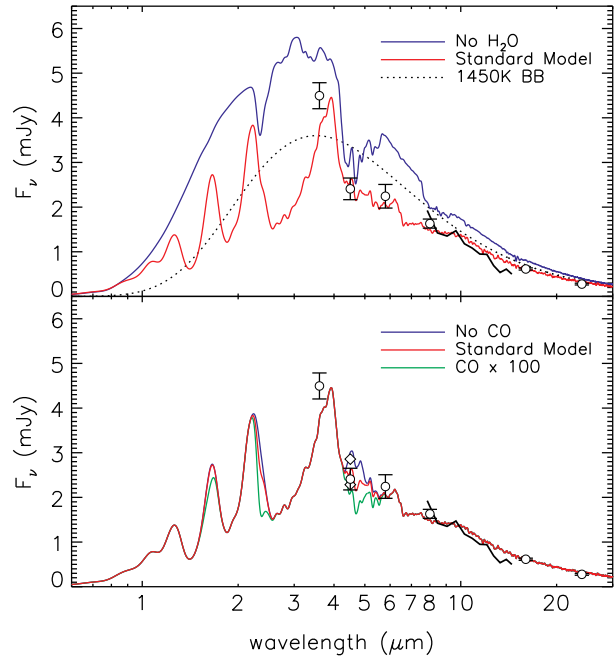}
\caption{{\em Top}: Synthetic spectrum of the solar abundance model shown in
Fig. \ref{epsfig}.  Observations \cite[]{Charb08} are shown with 1-$\sigma$
error-bars along with a 1450K black body (dotted line) and a model
with no water opacity (in blue).  {\em Bottom}: Standard model (red) compared
to models without CO (blue) and CO mole fraction equal to 100 times that of the
standard model. Diamonds are the band integrated model points at 4.5 $\micron$.
\label{ffig}}
\end{figure}

Interestingly, the $Spitzer$ IRAC data (3.6 to 8 $\mu$m) for HD 189733 show
evidence of H$_2$O and, to a lesser degree, CO absorption at the level
predicted by model atmospheres (the mole fractions of CO and H$_2$O for the
standard solar abundance model presented here are $6 \times 10^{-4}$ and $5
\times 10^{-4}$ respectively).  The IRAC colors in particular are consistent
with strong water absorption from $\sim$ 4 to 8 $\micron$. The good agreement
between the no-redistribution model and the observations (Figs. \ref{epsfig}
and \ref{ffig}) supports this interpretation.  Furthermore, the IRS spectrum
for HD 189733 reported to have no water absorption, is reasonably consistent
with the model everywhere except at the blue and red edges
\cite[]{FortenyMarely07}.  A flat $\epsilon_{\lambda}$ spectrum across 8
$\micron$ implies a very sharp rise in planet flux (see Fig.  \ref{ffig}) that
is inconsistent with the IRAC observations.

Spectra with different amounts of CO are compared to the observations in
the lower panel of Fig. \ref{ffig}.  An atmosphere absent of CO is excluded at
roughly 2-$\sigma$ and thus its presence is only tentatively suggested by the
data.  Unfortunately, the current precision of the 4.5 $\micron$ IRAC
photometry is not sufficient to test large departures of Carbon from solar
abundance or departures of CO from chemical equilibrium \cite[]{Cooper06}.

\section{An L-type Planet}

\begin{figure}
\hspace{-0.7cm}\includegraphics[width=9.5cm]{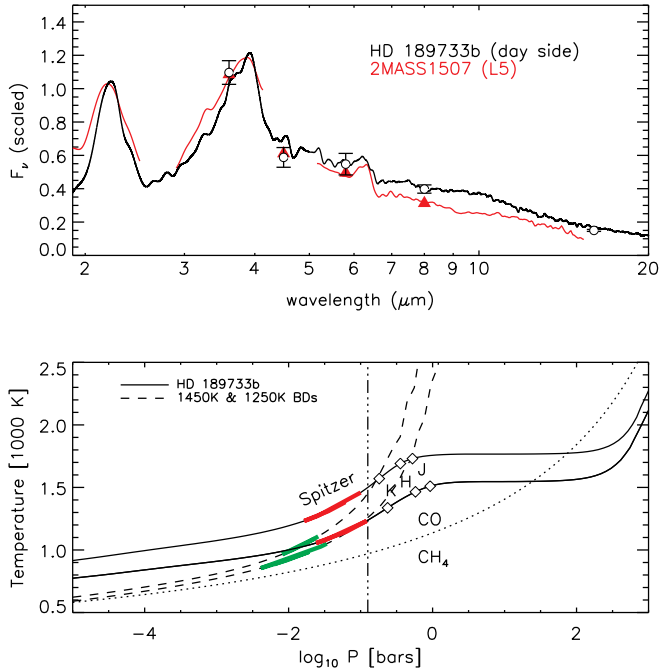}
\caption{ 
{\em Top}: The synthetic planet spectrum compared to a real IR spectrum of L5
brown dwarf 2MASS 1507. $Spitzer$ photometry for HD 189733b and 2MASS 1507 are
also shown, scaled to match at 3.6 $\mu$m.  
{\em Bottom}: Temperatures and pressures for no-redistribution (top solid line)
and full-redistribution (bottom solid line) models compared to 1250 and 1450K
brown dwarf models (dashed lines).  The $Spitzer$ and near-IR photospheric
depths are indicated by thick lines and filled symbols.  The vertical dash-dotted line is the
approximate boundary between efficient (higher P) and inefficient (lower P)
horizontal circulation.  The dotted curve denotes the region of equal CO/CH$_4$
equilibrium concentrations.
\label{bdfig}}
\end{figure}

Many hot-Jupiters have day side photospheric temperatures and pressures
comparable to those in L-type brown dwarfs and, despite potential differences
in formation and evolution, a great deal can be learned by comparing
observations of the former to observations of the latter.  Figure \ref{bdfig}
compares the HD 189733b observations and no-redistribution standard planet
model to IRTF and $Spitzer$ IR observations of an L5 brown dwarf, 2MASS 1507
\cite[]{Cushing06}, each scaled to match at 3.6 $\mu$m. The IRAC colors of the
planet and brown dwarf are nearly identical.  The decrease in flux across the
shortest two IRAC bands is a hallmark of CO and H$_2$O absorption in brown
dwarf atmospheres, lending support to the conclusions of the previous section.
The observed IRS spectrum of the brown dwarf is also similar to the planet
model spectrum, though a bit steeper.  Also, the $Spitzer$ photosphere in the
brown dwarf is at lower pressures than in the planet where the two temperature
structures have similar slopes and produce absorption features with similar
relative depths.
 
The spectral similarities between HD 189733b and 2MASS 1507 answers an
important question -- this planet's temperature structure can not be isothermal
at the photospheric depths probed by $Spitzer$.  This rules out one of the more
plausible explanations for why the first set of IRS spectra showed no observed
water absorption features.  Looking back at Fig. \ref{ffig}, one can see that
indeed the slope of the IRS planet spectrum, more or less, follows that of a
black body and, by itself, is very suggestive of a purely isothermal structure.
However, the 3.6 $\micron$ IRAC measurement, in concert with the longer
wavelength data, rules out any single black body spectrum at several $\sigma$.
Overall, the $> 3$ $\mu$m spectral region of the planet appears to be
remarkably similar to that of an L5 brown dwarf believed to have a solar
atmospheric composition dominated by water and a steep temperature structure.
However, as discussed below, irradiation driven atmospheric circulations most
likely cause the short-wavelength spectrum to depart significantly from that of
an L5 brown dwarf.

\section{Balancing the Budget}

Assuming HD 189733b is tidally locked, the incident stellar flux always enters
the planet through the same hemisphere.  This asymmetric heating may drive
strong atmospheric circulations that transport significant amounts of energy to
the night side \cite[]{Showman02, Cho03}.  Determining the efficiency of energy
redistribution across the day and night sides of a planet is, therefore,
important for understanding the global atmospheric circulation.  Recently
\cite{Knutson07b} measured the night side flux to be $\sim 64$ \% of the day
side flux at 8 $\mu$m. This suggests that some fraction of the absorbed
incident flux is, in fact, transported to the night side.  How then can the day
side $Spitzer$ observations so closely match the predicted flux from a model with
$zero$ horizontal energy transport, as shown above?

After a few 100 Myrs, internal energy leftover from the formation of HD 189733b
has mostly been lost, leaving the intrinsic luminosity $\sim 20,000$ times
smaller than what the planet receives from the star.  Thus, for old
short-period planets like HD 189733b, the star is the primary energy source.
Energy conservation requires that the planet's atmosphere radiate as much
energy as it receives and, if HD 189733b has a small albedo, the total
luminosity one should expect from the planet is determined entirely by the
stellar and orbital properties; ${\rm L}_{tot} \sim 2.6 \times 10^{-5}$
L$_\odot$.  Consequently, the true $bolometric$ luminosity of the planet's day
or night side (L$_{day}$ or L$_{night}$) is needed to precisely determine the
efficiency of redistribution.  Note that the efficiency of redistribution is
difficult to constrain with single band-pass photometry, since little leverage
on the bolometric luminosity is obtained, even with a complete phase curve.

\begin{figure}
\hspace{-0.6cm}\includegraphics[width=9.25cm]{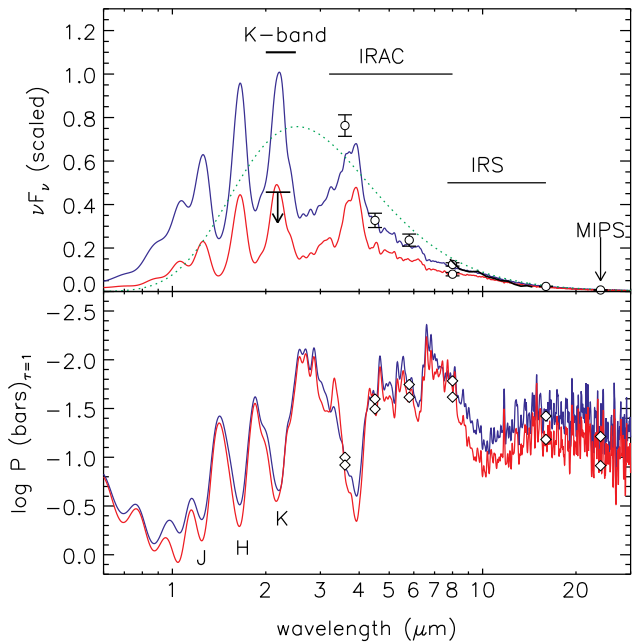}
\caption{{\em Top}: planet energy distributions assuming full-redistribution
(lower red line) and no-redistribution (top blue line) along with a 1450K
blackbody spectrum.  Over plotted are IR ground and space-based flux
measurements for HD 189733b (see text for references). The lower point at 8
$\mu$m is the night side flux measurement from \cite{Knutson07b}.  {\em
Bottom}: pressure at optical depth unity for both planet models.  The location
of the $Spitzer$ and near-IR band-integrated pressures are labeled.
\label{nufig}}
\end{figure}

Fig. \ref{nufig} shows $\nu F_{\nu}$ spectra for the full and no-redistribution
planet models.  Integrating the no-redistribution model spectrum (top curve)
for $\lambda >$ 3.2 $\mu$m accounts for 33\% of L$_{tot}$.  Constraining
L$_{day}$ further requires measurements at shorter wavelengths.  Fortunately,
recent ground based observations have provided an upper limit for the day side
$K$-band flux (less than 1.5 mJy)  that helps complete the energy balance
picture \cite[]{Barnes07}.  Though $\lambda_{p}^{max}$ $\sim 3.6$ $\mu$m, the peak
of the energy distribution sits at $K$-band making the $K$-band flux a very
useful gauge of the short-wavelength fraction of L$_{day}$.  By combining
the day side measurements, one can safely conclude that a sizable fraction of
the energy ($> 30$\%) is redistributed to the night side. Furthermore, these
data tell us that a 1-D no-redistribution model over estimates the flux at
$short$ wavelengths ($< 2.5$ $\mu$m).

The estimate of energy redistributed can be further improved by using the
\cite{Knutson07b} 8 $\mu$m night side measurement and the fact that only $\sim$
66\% of L$_{tot}$ remains to be shared across the entire night side spectrum
and the day side spectrum not covered by $Spitzer$.  The observed night side
flux is close to the full-redistribution model spectrum shown in Fig.
\ref{nufig} suggesting that the night side $Spitzer$ photospheric temperatures
are close to those predicted by the full-redistribution model.  Integrating
this spectrum (slightly scaled down to match the 8 $\mu$m night side
observation) across the complete wavelength range accounts for $\sim$ 43\% of
L$_{tot}$, leaving $\sim$ 24\% of L$_{tot}$ for the day side spectrum at
$\lambda < 2.5$ $\mu$m.  The day side $K$-band upper-limit is also close to the
full-redistribution model spectrum suggesting that both day and night side
near-IR spectra have similar flux levels.

As already discussed, the day side structure can not be entirely isothermal.
The $Spitzer$ data probe approximately the 0.1 to 0.01 bar region on the day
side while $J$, $H$, and $K$-bands probe higher pressures, $>$ 100 mbars (see
Fig. \ref{nufig}).  Lower fluxes on the day side at short wavelengths (as
indicated at $K$-band) compared to longer wavelengths suggests that energy
redistribution is highly depth-dependent and occurs fairly deep in the
atmosphere ($>$ 0.1 bars).  Similar depth-dependent redistribution has been
explored for HD 209458b and is the expected behavior in many hot-Jupiters
\cite[]{Cooper05,Seager05}.  Thus, the day and night sides probably have
similar temperatures at $P > 0.1$ bars which are close to the
full-redistribution structure shown in Fig.  \ref{bdfig}.  The steady supply of
incident flux from the star helps maintain the hot temperatures in the upper
layers of the day side while on the night side temperatures at $P< 0.1$ bars
are probably more similar to those of a cooler (T$_{eff} \sim 1250$K) brown
dwarf structure (see Fig.  \ref{bdfig}).  With lower temperatures, CH$_4$
begins to compete with CO as the dominant carbon bearing molecule on the night
side, especially in the upper atmosphere.  An important prediction of this
scenario is that one should $not$ expect significant phase variations at $J$,
$H$, and $K$-band for this planet.  Also, based on the models described here,
the 24 $\micron$ night side flux is predicted to be $\sim$ 75 \% of the day
side flux, a prediction that will soon be tested by $Spitzer$ observations.
A more rigorous constraint on redistribution will require additional
observations, especially of the night side, along with detailed global
circulation models.

\section{Conclusions}

HD 189733b has an atmosphere rich in water and, most likely, CO.  This planet's
atmosphere also has day side temperatures, pressures, and chemical composition
similar to a typical mid-L type brown dwarf.  The fraction of energy
redistributed from the day to night side has been estimated to be $\sim 43$\%
of the incident stellar flux.  Furthermore, the bulk of the energy transport
likely occurs at pressures higher than $\sim 0.1$ bars, indicating that the
energy redistribution mechanism is most efficient at high pressures.  It has
also been demonstrated that the photospheric depths probed by $Spitzer$ do not
reach the isothermal layers predicted by atmosphere models.  Consequently, an
isothermal structure can $not$ explain the flat $\epsilon_{\lambda}$ spectrum
recently measured by IRS.

\acknowledgements 
This paper benefited from the comments provided by the referee and many useful
discussions with Adam Showman, Brad Hansen, and Jonathan Fortney.   This
research was supported by NASA through Origins of Solar Systems (NNX07AG68) and
the $Spitzer$ theoretical research grants to Lowell Observatory.

\clearpage
\end{document}